\documentclass[twocolumn,showpacs,preprintnumbers,amsmath,amssymb]{revtex4}

\usepackage{graphicx}
\usepackage{dcolumn}
\usepackage{bm}

\begin{document}

\preprint{BAO-VA-07-2012}

\title{Extended Lorentz code of a superluminal particle}

\author{G Ter-Kazarian}

 \email{gago_50@yahoo.com}
\affiliation{
Byurakan Astrophysical Observatory\\
Byurakan 378433, Aragatsotn District, Armenia}


\begin{abstract}
While the OPERA experimental scrutiny is ongoing in the community,
in the present article we construct a toy model of {\it extended
Lorentz code} (ELC) of the uniform motion, which will be a well
established consistent and unique theoretical framework to explain
the apparent violations of the standard Lorentz code (SLC), the
possible manifestations of which arise in a similar way in all
particle sectors. We argue that in the ELC-framework the propagation
of the superluminal  particle, which implies the modified dispersion
relation, could be consistent with causality. Furthermore, in this
framework, we give a justification of forbiddance of
Vavilov-Cherenkov (VC)-radiation/or analog processes in vacuum. To
be consistent with the SN1987A and OPERA data, we identify the
neutrinos from SN1987A and the light as so-called {\it 1-th type}
particles carrying the {\it individual Lorentz motion code} with the
velocity of light $c_{1}\equiv c$ in vacuum as maximum attainable
velocity for all the $1$-th type particles. Thereby, we treat
superluminal muon neutrinos as so-called {\it 2-nd type} particles
carrying the individual Lorentz motion code with the velocity
$c_{2}$ as maximum attainable velocity for all the $2$-nd type
particles. For the  muon neutrinos mean energy $E_{\nu 2}=17.5$ GeV,
claimed velocity $(v_{\nu 2}-c)/c= 2.48\times 10^{-5}$, and expected
finite rest mass $m_{0}\approx 1eV/c^{2}$, we obtain then $
c_{2}/c\approx 17.5\times 10^{9}$.
\end{abstract}



\maketitle

\section{Introduction}
The special relativity (SR) encodes Lorentz symmetry as a particular
solution, so-called SLC, to make the interval defined in
four-dimensional Minkowski spacetime an invariant for the
transformation. This has as its crucial ingredient the universal
maximal velocity of light in vacuum\,-\, the limiting velocity in
Nature, remaining such for all the particles and in all inertial
frames of reference. Earlier studies of the neutrinos from the
supernova SN1987a, which is in the Large Magellanic Cloud at 51
kiloparsec from Earth,  bound scenarios of modification of neutrino
velocities. Due to the huge distance of the source of the neutrinos,
any small effect would therefore be largely amplified by the long
time of flight. These measurements set a stringent limit of $(v_{\nu
}-c)/c< 2 \times 10^{-9}$ for tens of MeV electron
neutrinos~\cite{Lon}. Consequently, the MINOS
collaboration~\cite{MI} (with 735 km baseline and a broad neutrino
energy spectrum peaked around 3 GeV) claimed the result
$\beta_{\nu}-1= (5.1 \pm 3.9) \times 10^{-5}, \quad \beta_{\nu}=
v_{\nu}/c$, which agrees at less than $1.4\sigma$ with the speed of
light. So it does not provide a strong evidence in favour of SLC
violating effects. However, the OPERA collaboration has claimed
instead a more precise result~\cite{OP}, which corresponds to a
$6\sigma$ effect for super-luminal propagation for muonic neutrinos,
thus confirming the MINOS results.  The data released are made of
16111 events detected in OPERA and correspond to about $10^{20}$
protons on target collected during the 2009, 2010 and 2011 CNGS
runs. The energetic muonic neutrinos $\nu_{\mu}$, mainly produced in
the decay, $\pi^{\pm}\rightarrow \mu^{\pm}+
\nu_{\mu}(\bar{\nu}_{\mu})$, cross the Earth's crust with mean
energy of $17.5$ GeV about $730$ km from CERN to the Gran Sasso at a
speed exceeding that of light $(v_{\nu }-c)/c= (2.48\pm
0.28(stat.)\pm 0.30(sys.))\times10^{-5}$ a significance of six
standard deviations. Shortly after this a second test is also
performed by using a beam with a short-bunch time - structure
allowing to measure the neutrino time of flight at the single
interaction level. The new analysis show consistent result with that
at the first version. We deliberately forebear from any presumption
of exotic hypothetical tachyonic or a pseudo-tachyonic behavior of
superluminal neutrinos, which seems nowhere near true even though if
one applies the Magorana's~\cite{Ma} additional solution of Dirac
equation with imaginary mass term. In this, a possible option for a
causal description is to allow tachyons to be incorporated in the
framework of absolute simultaneity for space-time events~\cite{S}
or, equivalently, the existence of a preferred reference frame but,
by its very existence, breaks Lorentz invariance. The dispersion
relation for tachyons, $E^{2}-c^{2}\vec{p}{}^{2}=-k^{2}c^{4}$, leads
to the large tachyonic mass of the OPERA muon neutrino: $k\approx
120$ MeV$/c^{2}$. This result is entirely incompatible with the last
quoted one determined from the kinematics of the pion decay at rest:
$\pi\rightarrow \mu\nu_{\mu}$~\cite{A}, which yielded
$m_{\nu}^{2}c^{4} = -0.016\pm0.023$ MeV$^{2}$~\cite{R}. Furthermore,
a tachyon may decay in number of exotic channels and so tachyonic
beam would be distorted upon its arrival at the Gran Sasso. However,
without ever referring to tachyonic physics, a hint to superluminal
neutrino propagation, if correct, already clashes with the SLC of
uniform motion of a particle at least for
the following two crucial reasons.\\
(1). A standard SR dispersion relation breaks down as the Lorentz
factor $\gamma= 1/\sqrt{1-\beta^{2}}$ becomes imaginary number and
the total energy cannot be
positive definite.\\
(2). A superluminal propagation violates the causality, so the CNGS
beam may become GSCN when seen from a sufficiently boosted reference
frame.

We begin by visualization of some properties of the violation of the
causality principle in the context of OPERA experiment. Suppose that
the measurement for both neutrinos and light involves two points on
the X axis of the reference frame S of the Earth. That is, the
superluminal muon neutrino beam is produced at the point A \,-\, by
accelerating protons to 400 GeV/c with the CERN Super Proton
Synchrotron(SPS), travels to the point B\,-\, OPERA detector at Gran
Sasso Laboratory, with the velocity $v_{\nu}$ and at B produces some
observable phenomenon, namely an identification of the $\tau-$lepton
created by its charged current (CC) interaction, which is the
signature of direct appearance mode in the $\nu_{\mu}\rightarrow
\nu_{\tau}$ channel. The starting of the neutrino travel at A and
the resulting phenomenon at B thus are being connected by the
relation of {\it cause} and {\it effect}. The time elapsing between
the cause and its effect as measured in OPERA experiment is
\begin{equation}
\begin{array}{l}
\Delta t=t_{B}-t_{A}=\frac{x_{B}-x_{A}}{v_{\nu}}=\frac{730
km}{v_{\nu}}\simeq 987.8 ns,
\end{array}
\label{R1}
\end{equation}
where $x_{A}$ and $x_{B}$ are the coordinates of the two points A
and B. Now in another system S', which is chosen so that the
Cartesian axes OX and O'X' lie along the same line and S' has the
velocity $V < c$ with respect to S, the time elapsing between cause
and effect would evidently be
\begin{equation}
\begin{array}{l}
\Delta
t'=t'_{B}-t'_{A}=\frac{t_{B}-\frac{V}{c^{2}}x_{B}}{\sqrt{1-\frac{V^{2}}{c^{2}}}}-
\frac{t_{A}-\frac{V}{c^{2}}x_{A}}{\sqrt{1-\frac{V^{2}}{c^{2}}}}=
\frac{1-\frac{v_{\nu}V}{c^{2}}}{\sqrt{1-\frac{V^{2}}{c^{2}}}}\Delta
t.
\end{array}
\label{R2}
\end{equation}
Suppose
\begin{equation}
\begin{array}{l}
V=c(1-1.48 \times10^{-5}),
\end{array}
\label{R4}
\end{equation}
then $\frac{v_{\nu}V}{c^{2}}$ would be greater than unity
\begin{equation}
\begin{array}{l}
\frac{v_{\nu}V}{c^{2}}\approx 1+ 1 \times10^{-5},
\end{array}
\label{R5}
\end{equation}
and hence, $\Delta t'$ becomes negative
\begin{equation}
\begin{array}{l}
\Delta t'\approx\frac{-1 \times10^{-5}}{\sqrt{1-(1-1.48
\times10^{-5})^{2}}}\Delta t\approx -1.82 <0.
\end{array}
\label{R6}
\end{equation}
In other words, {\it for an observer in system S' the effect which
occurs at  GS would precede in time its cause which originates at
CERN}. It is extremely hard to envisage a consistent theory having
such a logical impossibility. It is not excluded, however, that the
OPERA measurement report will eventually fail. Waiting for further
developments, see e.g.~\cite{Kl, BYY, CDP, AMR, Sv, Ko, Ha, LN, Gi,
AC, WWY,  Ale, LM, Io, ABP, Ga, Mat, CELZ, CNS, Tor, Sar, TL, GSS,
DV, Gu, Al, MS, DM, LW, PW, Co, AEM, KV, Gl, Das, MLM, CC, Alt, DR,
CFK}, nevertheless it is extremely challenging to explain SLC
apparent violations in a consistent theoretical framework. We have
proposed what is perhaps the minimal change in this regard. We
develop on the ELC of uniform motion which is a well established
consistent and unique theoretical framework to explain SLC apparent
violations. The possible manifestations of the latter arise in a
similar way in all particle sectors. It should be stressed that a
discovery of a superluminal particles would not invalidate the
Einstein's theory, as is notoriously claimed, but suggest an
extension in the superluminal sector. Since in the framework of ELC
a charged superluminal particle is allowed to propagate in vacuum
with a constant speed $v
> c$ higher than that of light, then at first sight it seems that
this particle will radiate VC-radiation until it is no longer
superluminal. So, this and analog processes are absent below a
characteristic energy and turn on abruptly once the threshold energy
is reached and, therefore, beam of superluminal particles would be
profoundly depleted as they propagate due to energy losses via the
VC-radiation /or analog processes. But as we will see, in the
ELC-framework the VC-radiation/ or analog processes of the
superluminal particle propagating in vacuum are forbidden. We will
proceed according to the following structure. In the next section,
we explain our idea of what is the individual Lorentz motion code of
a particle and lay a foundation of the ELC-framework. In this, a
modification of the dispersion relation for superluminal particle is
given. The causality principle for a superluminal propagation is
dealt with in section 3, in particular it was studied in the context
of OPERA experiment. In section 4, in the ELC-framework, we give a
justification of forbiddance of VC-radiation/or analog processes of
a superluminal particle propagating in vacuum. The concluding
remarks are presented in section 5.

\section{The ELC-framework}
The Lorentz transport equations are so constructed as to make the
interval between the infinitely close to each other two events
defined in Minkowski spacetime
\begin{equation}
\begin{array}{l}
d s^{2}=(d {x^{0}})^{2}-(d {x^{1}})^{2}-(d {x^{2}})^{2}-(d
{x^{3}})^{2}=inv,
\end{array}
\label{R7}
\end{equation}
an invariant for the transformation, and from the equality of the
infinitesimal intervals there follows the equality of the square of
the "length" of the radius four-vectors. Consider again two systems
S and S' in relative motion. A system S' is chosen so that the
Cartesian axes OX and O'X' lie along the same line, also denote
$(x^{1},x^{2},x^{3})\equiv (x, y, z)$. The Lorentz transformation
always retains its general form
\begin{equation}
\begin{array}{l}
x^{0'}=x^{0}ch \beta +x sh \beta, \quad x'=x ch \beta +x^{0} sh
\beta,\\
y'=y,\quad z'=z,\quad ch\beta=\frac{1}{\sqrt{1-\beta^{2}}},\quad
\beta\equiv\frac{x}{x^{0}}.
\end{array}
\label{R8}
\end{equation}
The standard SR theory encodes Lorentz symmetry as a {\it particular
solution (SLC-framework) to ~(\ref{R7})}:  namely introducing a
notion of 'time',  for all inertial frames of reference S, S',
S",..., we have then  SLC-relations:\phantom{a} $x^{0}= c t,\quad
x^{0'}= c t', \quad x^{0"}= c t",\dots$, agreed with ~(\ref{R8}).
Here, for further simplification we suppose that the starting-point
for 'time' measurements in the two systems is taken so that $t$ and
$t'$ are equal to zero when the two origins O and O' are in
coincidence. The SLC, in fact, is Einstein's postulate that the
velocity of light $(c)$ in free space appears the same to all
observers regardless the relative motion of the source of light and
the observer. Treating the SLC entirely as properties of
four-dimensional Minkowski space, it implies the velocity of light
$(c)$ to be universal maximum attainable velocity of a material body
found in this space. Even though, if for a moment we take the light
as a not interacting (unobservable) hidden physical state,
nevertheless the SLC will hold and, as before, the velocity of light
$(c)$ will be maximum attainable velocity for all other particles.
However, it is possible to preserve Lorentz covariance in a theory
also with a formal {\it general solution} to ~(\ref{R7}), which can
be obtained if the {\it 'time' at which event occurs is extended by
allowing an extra dependence on the 'different type' readings
$t_{i}$ $\,(i=1,2,...)$, which satisfy for all inertial frames of
reference S, S', S",...,  so-called ELC-relations:}
\begin{equation}
\begin{array}{l}
x^{0}\equiv c_{1}t_{1}=\dots=c_{i}t_{i}=\dots,\\
x^{0'}\equiv c_{1}t'_{1}=\dots=c_{i}t'_{i}=\dots,\\
\dots\dots\dots\dots\dots\dots\dots\dots\dots\dots
\end{array}
\label{R9}
\end{equation}
agreed with ~(\ref{R8}), where $c_{1}\equiv c$ is the speed of light
in vacuum, and $c_{i}> c_{1}, \,(i=2,3...)$ are speeds of the
additional 'light-like' states, higher than that of light. These
missing ingredients are the shortcoming of ELC-framework, which will
be motivated elsewhere. Phenomenologically such 'light-like' states
can be easily accommodated if for now to think of them as being the
hidden, or it may very well be that unobserved yet, states that
constitute the ELC. This assumption has the important consequence
that it may not be too unreasonable to link the existence of a
different type readings $t_{i}$  to the existence of a {\it
different type} of particles. This will call for a complete
reconsideration of our ideas of Lorentz motion code, to be now
referred to as the {\it individual code of a particle} \,-\,as its
intrinsic property. This observation allows us to lay forth a toy
model of the ELC, at which SLC violating new physics appears. That
is to say, {\it the $i$-th type particle in Minkowski space carries
an individual Lorentz motion code with its own maximum attainable
velocity $c_{i}$ ('its own velocity of light-like state').} The
clock reading $t_{i}$ can be used for the $i-$th type particle, the
velocity of which reads $v_{i}=x/t_{i}=c_{i}x/x^{0}$, so
$\beta=v_{1}/c_{1}=\dots v_{i}/c_{i}=\dots\equiv v/c= x/x^{0}$. If
$v_{i}=c_{i}$ then $v_{1}=c_{1}$, and the proper time of
'light-like' states are described by the null vectors $d
s^{2}_{1}=\dots d s^{2}_{i}=\dots=0$.  The extended Lorentz
transformation equations for given $i$-th and $j$-th type clock
readings can be written then in the form
\begin{equation}
\begin{array}{l}
x'=\gamma (x-vt),\quad t'_{i}=\gamma
\frac{c_{j}}{c_{i}}(t_{j}-\frac{v_{j}}{c_{j}^{2}}x), \\
y'=y,\quad z'=z.
\end{array}
\label{R10}
\end{equation}
Hence, like the standard SR theory, regardless the type of clock, a
metre stick travelling with system S measures shorter in the same
ratio, when the simultaneous positions of its ends are observed in
the other system S': $dx'=dx/\gamma$. Furthermore, a time interval
$dt_{i}$ specified by the $i-$th type readings, which occur at the
same point in system S ($dx=0$), will be specified with the $j-$th
type readings of system S' as $dt'_{j}=\gamma (c_{i}/c_{j})dt_{i}$.
Consequently, the modified transformation equation for velocity is
\begin{equation}
\begin{array}{l}
v_{i}=\frac{c_{i}}{c_{j}}
\frac{v'_{j}+V_{j}}{1+\frac{v'_{j}V_{j}}{c_{j}^{2}}}.
\end{array}
\label{R11}
\end{equation}
Here we have called attention to the fact that the mere composition
of velocities which are not themselves greater than that of $c_{i}$
will never lead to a speed that is greater than that of $c_{i}$.
Inevitably in the ELC-framework a specific task is arisen then to
distinguish the type of particles. This evidently cannot be done
when the velocity ranges of  different type particles intersect. To
reconcile this situation, we note that, according to (\ref{R9}), we
may freely interchange the types of particles in the intersection.
Therefore, we adopt following convention. With no loss of
generality, we may re-arrange a general solution that the particles
with velocities $v_{1}< c_{1}$, regardless their type, will be
treated as the 1-th type particles and, thus,  a common clock
reading for them and light will be set as $t_{1}$. This part of a
formalism  is completely equivalent to the SLC-framework.
Successively, the particles, other than 'light-like' ones, with
velocities in the range $c_{i-1}\leq v_{i}< c_{i}$, regardless their
type, will be treated as the i-th type particles and, thus, a common
clock reading for them and 'light-like' state $(i)$ will be set as
$t_{i}$. The invariant momentum
\begin{equation}
\begin{array}{l}
p^{2}_{i}=p_{\mu i}p^{\mu}_{i}=\left(\frac{E_{i}}{c_{i}}\right)^{2}-\vec{p}_{i}^{2}=m_{0\,i}^{2}c_{i}^{2}=\\
p^{2}_{1}=p_{\mu
1}p^{\mu}_{1}=\left(\frac{E_{1}}{c_{1}}\right)^{2}-\vec{p}_{1}^{2}=m_{0}^{2}c_{1}^{2},
\end{array}
\label{R12}
\end{equation}
introduces a {\it modified dispersion relation} for $i-$th type
particle:
\begin{equation}
\begin{array}{l}
E_{i}^{2}=\vec{p}_{i}^{2}c_{i}^{2}+m_{0
i}^{2}c_{i}^{4}=\vec{p}_{i}^{2}c_{i}^{2}+m_{0
1}^{2}c_{1}^{2}c_{i}^{2},
\end{array}
\label{DR12}
\end{equation}
where the mass of $i-$th type particle has the value $m_{0\,i}$,
when at rest,  the positive energy is
\begin{equation}
E_{i}=m_{i}c_{i}^{2}=\gamma m_{0 i}c_{i}^{2}=\gamma m_{0
1}c_{1}c_{i}, \label{DR13}
\end{equation}
and $\vec{p}_{i}=m_{i}\vec{v}_{i}=\gamma m_{0 i}\vec{v}_{i}$ is the
momentum. The relation~(\ref{DR13}) modifies the well-known
Einstein's equation that energy $E$ always has immediately
associated with it a positive mass $m_{i}=\gamma m_{0 i}$, when
moving with the velocity $\vec{v}_{i}$. Having set the theoretical
background, we now turn to discuss some consequences for the
superluminal propagation of particles. The next sections are devoted
mainly to these questions.

\section{The
causality principle for a superluminal particle} In the
ELC-framework of uniform motion,  the time elapsing between the
cause and its effect as measured  for the $i-$th type superluminal
particle is
\begin{equation}
\begin{array}{l}
\Delta t_{i}=t_{iB}-t_{iA}=\frac{x_{B}-x_{A}}{v_{i}}.
\end{array}
\label{R14}
\end{equation}
In another system S', which is chosen as before and has the
arbitrary velocity $V\equiv V_{j}$ with respect to S, the time
elapsing between cause and effect would be
\begin{equation}
\begin{array}{l}
\Delta
t'_{i}=\frac{1-\frac{V_{j}}{c_{j}}\frac{v_{i}}{c_{i}}}{\sqrt{1-\frac{V^{2}_{j}}{c_{j}^{2}}}}\Delta
t_{i}\geq 0,
\end{array}
\label{R16}
\end{equation}
where according to ~(\ref{R9}), $t_{iB}=(c_{j}/c_{i})t_{jB}$ and
$t_{iA}=(c_{j}/c_{i})t_{jA}$. That is, {\it the ELC-framework
recovers the causality for a superluminal propagation}, so the
starting of the superluminal impulse at A and the resulting
phenomenon at B are being connected by the relation of cause and
effect in arbitrary inertial frames. This completes the theoretical
discussion and we now turn to the SN1987A and OPERA experimental
inputs. To reconcile these data, we identify the {\it neutrinos from
SN1987A and the light as the 1-th type particles, and the muon
neutrinos as the 2-nd type superluminal particles}. While, as
discussed in the previous section, for the muon neutrinos mean
energy $E_{\nu 2}=17.5$ GeV and expected
 finite rest mass $m_{0 1}\simeq 1$eV$/c_{1}^{2}$ $\,\,(c_{1}\equiv c)$, the
 maximum attainable velocity $c_{2}$ for all the $2$-nd type
particles can be obtained as
\begin{equation}
\begin{array}{l}
c_{2}=\frac{c_{1}}{\gamma} \frac{E_{\nu
2}}{m_{01}c_{1}^{2}}=c_{1}\frac{E_{\nu
2}}{m_{01}c_{1}^{2}}\sqrt{1-\frac{v_{2}^{2}}{c_{2}^{2}}}
 \approx 17.5\times 10^{9}c.
\end{array}
\label{R19}
\end{equation}

\section{A forbiddance of VC-radiation/or analog processes
in vacuum} Assuming that the Lorentz invariance is violated
perturbatively in the context of conventional quantum field
theory~\cite{Hor, Vi,  MMS, Gl, Ale, AEM}, this can induce the muon
neutrino radiative decay $(\nu_{\mu}\rightarrow \nu_{\mu}+\gamma)$
and the three body decay - for example, the Z-strahlung radiation
$(\nu_{\mu}\rightarrow \nu_{\mu}+Z\rightarrow \nu_{\mu} +e^- +e^+)$,
otherwise forbidden but now kinematically permitted, being the
analog to VC-radiation~\cite{Gl, Das, MLM, CC, Alt, DR, CFK}. Such
processes will lead to the fast energy loss of neutrinos once the
threshold is reached. This implies that the Lorentz non-invariant
contribution added a shift in the momentum, which will result in a
shift in the maximum attainable velocity of the particle from the
velocity of light $c$, which remains the maximum attainable velocity
for all other particles except the muon neutrino. However, the
ELC-framework evidently forbids VC-radiation/or analog processes in
vacuum in a similar way in all particle sectors. Actually, in this
framework we have to set, for example, $k_{1}=(\frac{\omega}{
c_{1}}, \vec{k_{1}})$ for the 1-th type $\gamma_{1}$ photon,
provided $\vec{k_{1}}=\vec{e}_{k}\frac{\omega}{c_{1}}$, and
$p_{2}=(\frac{E_{2}}{ c_{2}}, \vec{p}_{2})$ for the 2-nd type
superluminal particle. Then the process  ($l_{2}\rightarrow
l_{2}+\gamma_{1}$) becomes kinematically permitted if and only if
\begin{equation}
\begin{array}{l} k_{1} p_{2}=\frac{\omega}{c_{1}}
\frac{E_{2}}{
c_{2}}\left(1-\vec{e}_{k}\frac{\vec{v}_{2}}{c_{2}}\right)= 0,
\end{array}
\label{R24}
\end{equation}
which yields  $\omega \equiv 0$ because of
$\left(1-\vec{e}_{k}\frac{\vec{v}_{\nu 2}}{c_{2}}\right)\neq 0$. So
in the case at hand, the {\it VC-radiation/or analog processes are
forbidden.} This evades constraints due to VC-like processes since
the superluminal neutrino $\nu_{\mu 2}$ does not actually travel
faster than the speed $c_{2}$, and that allows the arrival at the
Gran Sasso of superluminal neutrinos $\nu_{\mu 2}$ with the velocity
claimed by OPERA of any specific energy, without having lost of
their energies. In what follows, in ELC-framework we discuss the
VC-radiation  of the charged superluminal particle propagating in
vacuum with a constant speed $v_{2}> c_{1}$ higher than that of
light. Recall that, a charged particle ($e\neq 0$) moving in a
transparent, isotropic and non-magnetic medium with a constant
velocity higher than velocity of light in this medium is allowed to
radiate, so-called Vavilov-Cherenkov radiation. The energy loss per
frequency is~\cite{La}
\begin{equation}
\begin{array}{l}
d F= -d\omega \frac{ie^{2}}{2\pi}\sum
\omega\left(\frac{1}{c^{2}}-\frac{1}{\varepsilon
v^{2}}\right)\int\frac{d\zeta}{\zeta},
\end{array}
\label{R25}
\end{equation}
where the direction of the velocity $\vec{v}$ is chosen to be
$x-$direction: $k_{x}=k\cos \theta=\omega/v$, $\,k=n\omega/c$ is the
wave number $n=\sqrt{\varepsilon}$ is the real refractive index,
$\varepsilon$ is the permittivity. The summation is over terms with
$\omega=\pm|\omega|$, and a variable
\begin{equation}
\begin{array}{l}
\zeta=q^{2}-\omega^{2}\left(\frac{\varepsilon}{c^{2}}-\frac{1}{v^{2}}\right)
\end{array}
\label{R26}
\end{equation}
is introduced, provided $q=\sqrt{k_{y}^{2}+k_{z}^{2}}$. The
integrand in~(\ref{R25}) is strongly peaked near the singular point
$\zeta=0$, for which $q^{2}+k_{x}^{2}=k^{2}$.  Using standard
technique, the  formula (\ref{R25}) can  be easily further
transformed to be applicable in ELC-framework for the charged
superluminal particle of $2$-nd type propagating in vacuum (i.e. if
$\varepsilon=1$) with a constant speed $v_{2}$ higher than that of
light ($c_{1}\leq v_{2}< c_{2})$:
\begin{equation}
\begin{array}{l}
d F= -d\omega \frac{ie^{2}}{2\pi}\sum
\omega\left(\frac{1}{c^{2}_{2}}-\frac{1}{v^{2}_{2}}\right)\int\frac{d\zeta}{\zeta},
\end{array}
\label{R27}
\end{equation}
and, respectively, (\ref{R26}) becomes
\begin{equation}
\begin{array}{l}
\zeta=q^{2}_{1}-\omega^{2}\left(\frac{1}{c^{2}_{2}}-\frac{1}{v^{2}_{2}}\right),
\end{array}
\label{R28}
\end{equation}
where $q_{1}=\sqrt{k_{y1}^{2}+k_{z1}^{2}}$, $\,q^{2}_{1}+k_{x1}^{2}=
k_{1}^{2}=\omega^{2}/c_{1}^{2}$, and now $k_{x1}v_{2}=\omega$. We
have then
\begin{equation}
\begin{array}{l}
\zeta=\frac{\omega^{2}}{c^{2}_{2}}\left(
\frac{c_{2}^{2}}{v^{2}_{2}\cos^{2}\theta}-1\right)\neq 0,
\end{array}
\label{R29}
\end{equation}
because of $v_{2}< c_{2}$, and that the integral~(\ref{R27}) is
zero, since the integrand has no poles. Hence, as expected, the {\it
VC-radiation of a charged superluminal  particle as it propagates in
vacuum is forbidden}.

\section{Concluding remarks}
To summarize, we have pointed out that if there were some
superluminal particles which indeed violet the SLC, the ELC will be
a well established consistent and unique theoretical framework to
explain these apparent violations, the possible manifestations of
which arise in a similar way in all particle sectors. We drastically
change our ideas of Lorentz motion code, treating it as an {\it
individual code of a particle} \,-\,as its intrinsic property. The
shortcoming of the ELC-framework, of course, are the missing
ingredients of the heuristic 'light-like' states other than that of
light, which will be motivated elsewhere. We for now think of them
as being the hidden, or it may very well be that of unobserved yet,
states that constitute the ELC. The ELC-framework recovers the
dispersion relation and the causality for a superluminal
propagation, as well as forbids VC-radiation/or analog processes of
a superluminal particle propagating in vacuum. In the context of the
OPERA experiment, we treat superluminal muon neutrinos as  2-nd type
particles carrying the individual Lorentz motion code with the
velocity $c_{2}$ as maximum attainable velocity for all the 2-nd
type particles. We obtain then $c_{2}\approx v_{\nu 2}(1+1.65\times
10^{-21})$.





\end{document}